\documentclass[aps, amssymb, amsmath, reprint]{revtex4-1} 

\usepackage{graphicx}  % needed for figures
\usepackage{dcolumn}   % needed for some tables
\usepackage{physics}
\usepackage{color}
\usepackage{appendix}
\usepackage{ulem}
\usepackage{soul}
\definecolor{orange}{rgb}{0.8, 0.3, 0}

\definecolor{blueviolet}{rgb}{0.2, 0.2, 0.6}
\usepackage[pdftex, % FOR PDFLATEX ONLY
bookmarks=true, % Bookmark bar
colorlinks=true,
allcolors=blueviolet,
pdfstartview={FitH}, % FitBH
]{hyperref}

\newcommand{\Er}{\text{Er}^{3+}}
\newcommand{\ktot}{\kappa_\text{tot}}

\newcommand{\ginh}{\Gamma_\text{inh} } 
\newcommand{\crystal}{\text{CaWO}_{4} }

\begin{document}

\title{Pulse-tube-vibration induced Larmor frequency jitter in anisotropic Kramers spins }

\author{R.~Shaik$^{1}$}
\author{P.~Kesarwani~$^{1}$}
\author{S.~Rajendran$^{1}$}
\author{S.~E.~Kubatkin$^2$}
\author{A.~V.~Danilov$^2$}
\author{S.~E.~de~Graaf$^{3}$}
\author{V.~Ranjan$^{1}$}
\email{vranjan@tifrh.res.in}

\affiliation{$^1$ Tata Institute of Fundamental Research, Hyderabad 500046, India}
\affiliation{$^2$Department of Microtechnology and Nanoscience MC2, Chalmers University of Technology, SE-41296 Goteborg, Sweden}
\affiliation{$^3$National Physical Laboratory, Teddington TW11 0LW, United Kingdom }

%% Abstract #######################################
\begin{abstract} %% should be less than 600 characters incl spaces
We investigate the role of pulse-tube mechanical vibrations towards spin frequency noise in a cryogen-free dilution refrigerator. Our system comprises of a low mode-volume superconducting resonator coupled to 50 ppm $\Er:\text{CaWO}_4$ spins with highly anisotropic g-factors. By measuring single-shot spin echoes using narrow-bandwidth three-pulse sequences, we directly observe a vibration induced stochastic signal resulting from the spin's Larmor frequency moving into and out of the detection bandwidth. The sensitivity to these vibrations is substantially suppressed when a) the external magnetic field is aligned along the symmetry c-axis and b) spin temperature is modestly increased to 60~mK. We also benchmark the system as a quantum sensor for angular fluctuations, extracting a noise floor of 350 nanodegrees.
\end{abstract}

\maketitle
%%##################### Introduction #############################
\section{Introduction}
Electron spin dephasing is primarily driven by microscopic magnetic field fluctuations in their surrounding environment, e.g. arising from mutual dipolar flip-flops of neighboring nuclear and electron spins, and coupling to the lattice phonon modes~\cite{schweiger_principles_2001}. At sub-kelvin temperatures, however, dominant flip-flops of electron spins get suppressed, and spin coherence times can potentially be limited by macroscopic factors tied to the specific measurement setup. One such factor is mechanical vibration of the sample, stemming from inadequate vibration-isolation from sources like vacuum pumps and pulse-tube (PT) compressors~\cite{kalra_vibration-induced_2016, britton_vibration-induced_2016, louchet-chauvet_piezospectroscopic_2019, kono_mechanically_2024}. The isolation is especially challenging for cryogen-free dilution refrigerator systems in which the magnet (4~K) and samples (10~mK) are needed to be maintained at different temperatures, preventing the use of mechanical links~\cite{caparrelli_vibration-free_2006}. 

The increasing availability of magnet-integrated, cryogen-free dilution refrigerators has opened new avenues for conveniently investigating the interactions between spin systems and their environment~\cite{budakian_roadmap_2024, Stern2026, goldfarb_epr_2026}. In particular, use of superconducting circuits with low-mode-volume has led to unprecedented measurement sensitivities while preserving the standard coherent inductive detection techniques~\cite{ranjan_electron_2020,budoyo_electron_2020}. More recently, spectroscopy of single proximal nuclear spins with Hertz spectral resolution~\cite{travesedo_precision_2026} has been demonstrated via microwave photon counting~\cite{Lescanne2020} at millikelvin temperatures. Additionally, electron spin coherence times now routinely reach orders of tens of milliseconds to seconds, promising their use as solid-state quantum memories~\cite{lvovsky_optical_2009,morton_storing_2018}. Understanding spin dephasing from slow drifts due to macroscopic sources is therefore essential for developing their practical use in quantum computing and sensing. 

In this work, we report on spin dephasing due to angular fluctuation of the electron spins quantization axis~\cite{kumar_strongly_2026} caused by the mechanical action of the pulse-tube compressor [Fig.~\ref{fig1:Schematics}(a)]. To this end, we perform measurements on $\Er$ ions in tetragonal $\crystal$ crystal possessing strongly anisotropic g-factors~\cite{antipin_paramagnetic_1968}, and due to which their Larmor frequency becomes highly sensitive to even small angular fluctuation of the magnetic field direction even if minimal field-inhomogeneities exist across the small resonator mode-volume. We are able to directly observe the effect of vibration in the amplitude fluctuations of single-shot spin echoes acquired using three-pulse inversion recovery sequences for the cases of PT running and switched off [Fig.~\ref{fig1:Schematics}(b)]. Such fluctuations in echo magnitude are however strongly suppressed when magnetic field is aligned with at the symmetry $c$-axis, even though the PT is kept on. These are the primary results of this work, and which we discuss further below.  

%################## Setup
\section{Measurement setup}

%Because of bulk distribution of Er$^{3+}$ and narrow inductor width $1~\mu$m, control pulses create extremely inhomogeneous tipping angles for different spins. 

The electron spin system under study here is the nuclear-spin free isotopes (78\% abundance) of 50~ppm Er$^{3+}$ ions in a tetragonal CaWO$_4$ crystal. This has been purchased from Surface Net Germany. The ESR measurements are performed with a superconducting resonator of frequency $\omega_0/2\pi = 4.43$~GHz and bare loss rate  of $\kappa/2\pi = 2.5 \pm 0.1~$MHz. The resonator couples to spins via a straight inductor of length $500~\mu$m (along $z-$axis, nearly parallel to the $c-$axis of the crystal) inside a cylindrical mode-volume volume of a radius $< 10~\mu$m. The magnetic field $B_\text{mag}$ is applied in the $ac$ plane of the crystal making an angle $\theta$ with c-axis. The square control pulses have durations of $1-4~\mu$s resulting in narrow bandwidth excitations and detection $\ll \kappa, \ginh$, where $\ginh$ is the inhomogeneous spin linewidth, FWHM here. More details of the setup can be found in Ref.~\cite{ranjan_spin-echo_2022}.

%Their nominal doping of 50 ppm corresponds to an ion concentration of $6.4 \times 10^{17} \text{cm}^{-3}$.

\begin{figure}[t!]
    \centering
\includegraphics[width=\columnwidth]{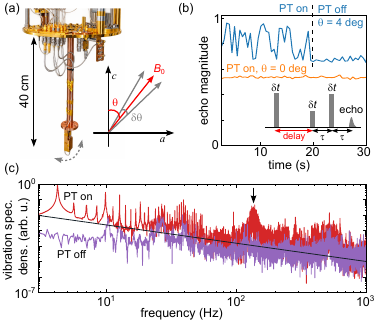} 
\caption{Pulse tube induced vibration of sample. (a) Sample assembly showing the crystal's change in angle with respect to applied magnetic field. The angle $\theta$ is with respect to $c$-axis of the crystal and varied in the $ac$ plane. (b) Repetitions of single shot echo magnitude acquired at two $\theta$ values. The cases of Pulse tube kept on and off are accordingly marked. The corresponding three-pulse inversion recovery sequence is shown in the inset. (c) Radial vibration amplitude spectral density acquired with an accelerometer at room temperature for cases of pulse tube on and off. Straight line showing $1/f$ dependence is a guide to the eye.}
    \label{fig1:Schematics}
\end{figure}
 
The crystal is glued on top of the resonator using vacuum grease (Dupont) and the sample assembly mounted on a 40~cm long probe-arm hanging form the mixing chamber plate into the bore of a vector magnet (their three axis referred to as $\{ z,~x,~y\}$-axis magnets). The magnet is anchored to the 4-K flange of the Bluefors LD250 dilution refrigerator. The sample box assembly consists of low permeability materials such as OFHC and brass to minimize field gradients. The silicone type grease transitions into a solid glassy phase~\cite{streque_low-temperature_2018} at low temperatures such that pulse tube vibrations are less likely to cause any sliding between the resonator and the crystal. However, the probe-arm can vibrate relative to the magnet resulting in a stochastic changes to spins' Larmor frequency during their detection. In our case, The nominal magnetic field inhomogeneity is around 0.1\% and 1\% over a $1~\text{cm}$ diameter of spherical volume for a magnetic field along the $z$-and $x-$ axis, respectively. 

The Radial vibrational amplitude spectral density measured using an accelerometer at the sample position and ambient conditions is plotted in Fig.~\ref{fig1:Schematics}(c). Their typical $1/f-$type decrease with frequency~\cite{chijioke_vibration_2010} show several orders of magnitude larger response for frequencies $<500$~Hz in the case of pulse tube turned on than that for switched off. On top of the periodic peaks separated by 1.4~Hz corresponding to the pulse-tube action, we also observe local broader peaks due to natural mechanical resonances of the cryostat assembly (simulating this is beyond the scope of present work). For example, there is one near 100~Hz indicated by an arrow in Fig.~\ref{fig1:Schematics}(c), which affects our measurements visibly as discussed further below. We note that due to space constraints in the magnet bore, the accelerometer measurements are possible only after removing the magnet. 

% ############### spin system
\section{Anisotropic electron spin frequency and linewidth}

Following the work by Mims in Ref.~\cite{mims_electric_1965}, the static Hamiltonian of $\Er$ electron spins (effective $S=1/2,~I=0$) in $\crystal$ in presence of magnetic \textbf{B} and electric fields \textbf{E} can be described by 

\begin{equation}
    H_0 = \mu_\text{B} \sum_{ij} g_{ij} B_i S_j + \sum_{ijk} T_{kij} B_i S_j E_k
     \label{Eq:hamiltonian}
\end{equation}

\noindent where $\mu_\text{B}$ is the Bohr-magneton, and $g_{ij}$ and $T_{kij}$ are elements of second rank \textbf{g}-factor tensor and third rank magneto-electric coupling tensor \textbf{T}, respectively. The latter can be constructed using a third rank tensor \textbf{d}, which has the same symmetries as a converse-piezo-electric tensor ~\cite{mims_electric_1965, mims_broadening_1966}. In our case, there is no external electric field, however, due to lack of inversion symmetry both intrinsic charge defects and defects introduced to compensate the trivalent ions of erbium substituting the divalent calcium ions do give rise to inhomogeneous electric fields. The second term in Eq.~\ref{Eq:hamiltonian} is nonetheless relatively much smaller and can be approximated as a linewidth broadening correction. Due to $S_4$ symmetry of the crystal~\cite{hempstead_paramagnetic_1960}, simplifications arise in both tensors. Taking crystal axes as the basis directions, firstly the \textbf{g}-factor becomes diagonal with values $g_{a}= g_{b} = 8.38,~g_{c}=1.247$. The angular dependence of the spin transition frequency at a given magnetic field $B_\text{mag}$ then simply reads as 
\begin{equation}
    \nu = \frac{\mu_\text{B}}{h} B_\text{mag} \left[ (g_c \cos (\theta) )^2 + (g_a \sin (\theta) )^2  \right]^{1/2}.
    \label{eq:spinfrequency}
\end{equation}

\noindent Secondly, the tensor elements $d_{kij}$ can also be described in the Voigt notation as a $3 \times 6$ matrix with only four unique and non-zero elements,

\begin{equation}
d _{kij} \equiv \left(
\begin{array}{ccc|ccc}
0 & 0 & 0 & d_{14} & d_{15} & 0 \\
0 & 0 & 0 & -d_{15} & d_{14} & 0 \\
d_{31} & -d_{31} & 0 & 0 & 0 & d_{36}
\end{array}
\right),
\end{equation}
where $\{ d_{14}, d_{15}, d_{31} , d_{36} \} = \{ 1.06,~-1.47,~-9.7,~5.2\}\times 10^{-6} ~\text{cm/V}$~\cite{mims_electric_1965}, from which changes in g-factor follow~\cite{mims_electric_1965}

\begin{equation}
\begin{split}
\delta \text{g}^2 = \sum_k E_k ( d_{k1} l^2 & +  d_{k2} m^2 + d_{k3} n^2   \\
                                            & + 2d_{k4} mn + 2d_{k5} ln + 2d_{k6} lm ).
\end{split}
\end{equation}
\noindent The factors $l, ~m, ~n$ are direction cosines of a unit vector parallel to the magnetic field. In particular, when magnetic field is applied in the $ab$ plane, only the component of the electric field along the c-axis contributes to change in the spin's Larmor frequency such that spin linewidth $\Delta \omega = \left| \partial \omega/ \partial E_c \right| \Delta E_c$ as first demonstrated at a temperature of 4.2~K in Ref.~\cite{mims_broadening_1966} and recently reproduced in Ref.~\cite{dantec_twenty-threemillisecond_2021} at millikelvin temperatures using a setup similar to the present work. For the general case $\Delta \omega$ is given as~\cite{scott_paramagnetic_1965} 

\begin{equation}
(\Delta \omega)^2 = \left( \frac{\partial \omega}{\partial E_a}\Delta E_a
\right)^2 + \left( \frac{\partial \omega}{\partial E_b}\Delta E_b
\right)^2 + \left( \frac{\partial \omega}{\partial E_c}\Delta E_c \right)^2
.
    \label{Eq:spinlinewidth}
\end{equation}

\begin{figure}[t!]
    \centering
\includegraphics[width=\columnwidth]{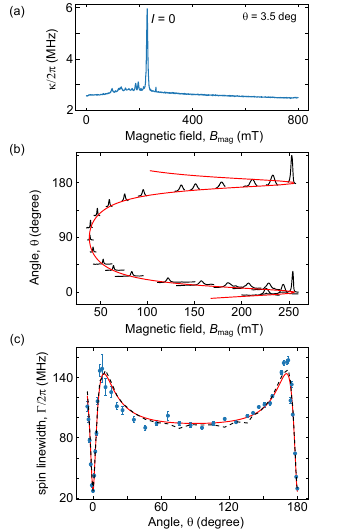} 
\caption{Anisotropic spin properties. (a) cw spectra, resonator linewidth $\ktot$ vs magnetic field magnitude $B_\text{mag}$, acquired at several angles $\theta$. The theoretical positions of resonant field values is also plotted. Spectra have vertical offsets proportional to their respective $\theta$(In the crystal $ac$ plane). (b) Angular dependence of the extracted inhomogeneous spin linewidth $\ginh$ (symbols: measurement, solid line: fit to theory, dashed line: Monte Carlo simulation). Error bars represent the standard deviations associated with extraction. }
    \label{fig1:Spectra}
\end{figure}

The highly anisotropic spin spectra acquired using continuous wave VNA measurements and respective extracted spin linewidth $\ginh$ (FWHM) at different angles $\theta$ are plotted in Fig.~\ref{fig1:Spectra}. The corresponding fits (solid lines) to Eq.~\ref{eq:spinfrequency} and Eq.~\ref{Eq:spinlinewidth} display an excellent match to theory. To minimize the number of free parameters in the fit to spin linewidth, we have assumed equal electric field strengths across axes yielding $\Delta E_a = \Delta E_b = \Delta E_c \approx 340 \pm 16~$kV/cm. We also perform a Monte Carlo simulation using a 3D Holtsmark distribution ~\cite{holtsmark_uber_1919} of the charge defects with a probability density of

\begin{equation}
    P(E) dE = \frac{2E}{\pi E_0^2}dE \int_0^{\infty} x \sin \left( xE/E_0\right) \text{exp}{(-x^{3/2 } )} dx,
\end{equation}
where $x$ is the variable of integration. The characteristic electric field  $E_0 = \frac{q}{2 \epsilon_0 \epsilon_r} (4 n_d/15)^{2/3}$~\cite{romanovsky_electric_2005} sets the most probable value at $1.6E_0$ with a FWHM $\sim 2.2E_0$. The best fit is shown in  Fig.~\ref{fig1:Spectra}(c) as a dashed line yielding $E_0 = 130 \pm 15 $~kV/cm and doubly charged $q=2e$ defects density (e.g. missing Ca$^{2+}$ or O$^{2-}$ ions ) of $n_d \approx 2.5 \times 10^{18}~\text{cm}^{-3}$. This is roughly four times the nominal concentration of $\Er$ ions suggesting most defects are native to the crystal growth and not doping-related~\cite{choi_native_2026}. Similar defect density $n_d \approx 1.4 \times 10^{18}~\text{cm}^{-3}$ has been estimated in Ref.~\cite{mims_electric_1965} in a 40~ppm doped crystal. We have also used a baseline in the fit to Eq.~\ref{Eq:spinlinewidth}, which we tentatively attribute to a combination of natural misalignment of the crystal and strain-mediated line broadening from point defects, such as the site vacancies mentioned above~\cite{mims_electric_1965}.    

%% ########################################## Echo variances
\section{Vibration induced Larmor frequency jitter}
\begin{figure*}[t!]
    \centering
\includegraphics[width=\textwidth]{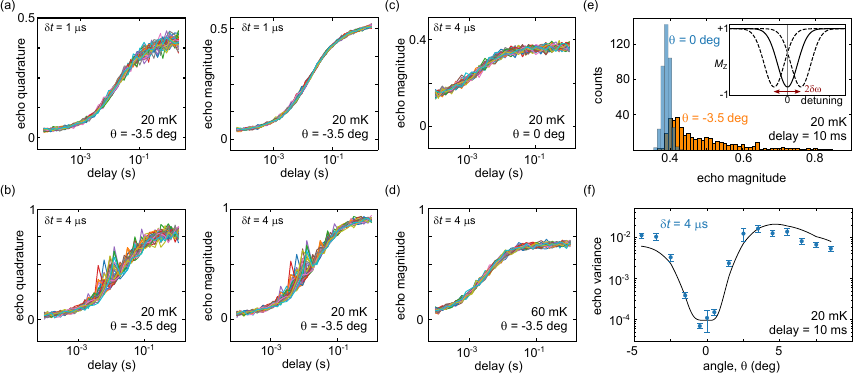} 
\caption{Vibration induced Larmor frequency jitter. (a-d) Single shot acquisitions of echo inversion recovery at different angles $\theta$ and temperatures as mentioned in the panels. There are 40 traces in each panel. The echo amplitudes are different across panels due to different loss rates of the resonator $\kappa_\text{tot}$ at spin resonances leading to a variable power transferred. (e) Echo magnitude histograms at $\theta = 0~\text{deg}$ and $\theta = -3.5~ \text{deg}$ at 20mK and a delay of 10ms. The inset shows a schematic of how mechanical vibration shifts the spectral hole (dashed gaussians) w.r.t the detection window (solid gaussian). (f) Echo magnitude variances (measured: symbols, model: line) at the delay of 10~ms for different angles at 20~mK. A 2$\sigma$ confidence interval was found using bootstrapping\cite{breiman2001statistical} on the 400 shot dataset to set error bars on the calculated variance.}
    \label{fig1:invEcho}
\end{figure*}

We measure spin echoes from three-pulse inversion recovery sequences $\pi - delay-\pi/2-\tau-\pi-\tau-echo$ [inset of Fig.~\ref{fig1:Schematics}(b)] to track the instantaneous changes in the spin's frequency within the narrow detection bandwidth roughly set by $\sigma_D \approx 1/\delta \tau $. The inversion pulse burns a spectral hole with a linewidth $\ll \ginh$, and whose central frequency fluctuates due to the mechanical vibration of the crystal axis. Subpopulation jitter within the detection window is then coherently mapped onto the fluctuating echo amplitudes via the subsequent 2-pulse Hahn echoes. The repetition time of the measurement sequences across all parameters is fixed at 5~s $>T_\text{1D}$, their direct spin-phonon relaxation times~\cite{Rajendran2026}. Since the $T_\text{1D}$ increases following $B_0^2$, which necessitates a slower repetition rate, the measurements have been limited to $|\theta| < 10~$deg. 

Single-shot echo quadrature amplitude at $\theta = -3.5~$deg acquired with pulse duration of $1~\mu$s (40 traces, second quadrature not shown) and delay range $100~\mu\text{s}-1~\text{s}$ are plotted in Fig.~\ref{fig1:invEcho}(a), left panel. The traces show clear amplitude fluctuations from slow phase noise drifts increasing at larger delays. The magnitude of the same echoes are devoid of such fluctuations [right panel of Fig.~\ref{fig1:invEcho}(a)], indicating associated frequency changes to be $\ll \Omega_\text{det}$. 
On the other hand, same measurements acquired with pulse durations $\delta t = 4~\mu$s show strong amplitude fluctuations even in the echo magnitudes [right panel of Fig.~\ref{fig1:invEcho}(b)]. Especially, we notice enhanced amplitude fluctuations near delays of 10~ms, which we will use below to study dependencies. Such fluctuations get suppressed near the c-axis ($\theta = 0$) and with modest increase of temperature to $\sim 60~$mK [Fig.~\ref{fig1:invEcho}(c,~d)]. We note that due to strong inhomogeneity of Rabi frequencies caused by our narrow inductors ($1~\mu$m), the average spin inversion is smaller than $50\%$.

To quantify the aforementioned changes, we have measured variances of echo magnitudes (400 single-shot acquisitions each) at a fixed delay of 10~ms and repetition time of 5~s at different angles. Figure~\ref{fig1:invEcho}(e) shows the histograms for two angles, $\theta = 0~\text{deg}$ and $\theta = -3.5~ \text{deg}$ clearly showing large differences in variance in the datasets. That a normal distribution is not observed makes sense since measured signal increases irrespective of positive or negative angular fluctuations. Full dataset of echo variance against angles are plotted in Figure~\ref{fig1:invEcho}(f) showing a symmetric response around 0 degree. We note that two angles $\theta = \mp 3.5~$deg are different in the lab frame $\theta_\text{lab} = \{0, 7\}~$deg due to slight misalignment of the crystal's $c$-axis with the $z-$axis magnet. The symmetric response in echo variance versus angle thus confirms that the additional field inhomogeneity from the second y-axis magnet does not contribute significantly towards the enhanced fluctuations of echo amplitude away from the $c$-axis. 

\begin{figure}[t!]
    \centering
\includegraphics[width=\columnwidth]{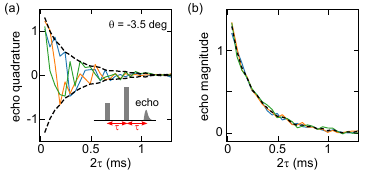} 
\caption{Two-pulse echo decay at $\theta = -3.5$deg. (a) Three traces of echo quadrature and (b) echo magnitude versus inter-pulse delay. Dashed lines are ($\pm$) averaged echo magnitudes capturing the envelope in the panel (a).}
    \label{fig1:2PE}
\end{figure}

For simplicity, we model the measured variance assuming a gaussian spectral hole of linewidth $\sigma_H$ due to the inversion $\pi$ pulse and a gaussian detection window due to the subsequent $\pi/2$ and $\pi$ pulses. The observed echo signal can be written as 
\begin{align}
    S(\delta \omega, ~\theta, ~T)  &= A(\theta,~T)\left[1- K(\theta,~T) F(\delta \omega) \right]
    \label{eq:signal}
\end{align}
where $K$ is experimentally determined constant containing the efficiency of the inversion pulse and subsequent $T_1$ decay. $F(\delta \omega) = \exp \left(- \frac{\delta \omega^2}{2 \left(\sigma^2_H + \sigma^2_D \right)} \right)$ is the filter function obtained by convolution of the spectral hole and the detection window as derived in the Appendix \ref{app:a}. Here, $\sigma_\text{eff} = \sqrt{\sigma^2_H + \sigma^2_D}$ is the effective bandwidth. In our case, we use square pulses with pulse duration $\delta t$ both during inversion and detection, $\sigma_\text{eff} \approx 2\pi \frac{\sqrt{2}}{\delta \text{t}}$. The factor A depends on the loss rate $\kappa$ of the resonator, temperature dependent polarization, the inhomogeneous linewidth discussed above, $T_2(\theta, ~T)$, instrumental amplification factors and is experimentally found using a Hahn echo sequence. Since the pulse tube action is not synced with the repetitions of the measurements, the noise can be treated random, and we take a Gaussian probability distribution for the detuning with a bandwidth, $\sigma_{\omega} = 2 \pi \left| \frac{\partial \nu}{\partial \theta} \right| \sigma_{\theta}$, calculated at each angle using Eq~\eqref{eq:spinfrequency}. Finally, the corresponding variance in echo magnitude is (see Appendix.~\ref{app:a})
\begin{align}
    \begin{split}
    \langle (\Delta S)^2 \rangle = A^2(\theta,~T) K^2(\theta,~T) &\bigl[  (1 +2 \Sigma^2(\theta, T))^{-0.5} 
    \\
                                                & - (1+ \Sigma^2(\theta, T))^{-1} \bigr]
   \end{split}
   \label{vareq}
\end{align}
where $\Sigma(\theta,~T) = \sigma_{\omega}/\sigma_\text{eff}$. Figure~\ref{fig1:invEcho}(f) shows the corresponding fit to the data using Eq.~\eqref{vareq} from which we extract $\sigma_{\theta} = 5 \times 10^{-4} \pm 1.8 \times 10^{-5}$ degree, translating to an average radial displacement of $L \sigma_\theta \sim $4.4~nm at delay of 10~ms. Here, the crystal length is $L=0.5~$mm. The extracted displacement value is an order of magnitude larger than the factory-tested value for our dilution fridge, which can be possibly related to the local maxima in the vibration spectral density near 100~Hz [Fig.~\ref{fig1:Schematics}(c)], but can also suggest presence of site- and installation-specific detrimental vibrational effects. One difference in our setup is the probe-arm of length 40~cm, hanging from the lowest plate of the dilution fridge, on which sample is mounted~[Fig.~\ref{fig1:Schematics}(a)]. Consequently, the large Larmor-frequency jitters lead to dramatic reduction of the spin coherence times, as can be readily observed in the single-shot quadrature measurements of two-pulse echoes [Fig.~\ref{fig1:2PE}], with dynamic phase shifts $\Delta \phi \approx \sigma_\omega \tau$ of order $\pi$ already starting to appear at short delays $2\tau = 200~\mu$s. At this shorter delays, we can also deduce a $\sigma_\theta\sim 28 \times 10^{-6}$ degrees, and an average radial displacement of $0.24$~nm.

While the observed angular jitter is a detrimental to spin coherence, the setup can conversely be used a quantum sensor for angular fluctuations~\cite{kumar_strongly_2026}. To this end, we acquire standard deviations of echo phase with two-pulse sequences at short delays $\tau = 20~\mu$s and $\theta =0$ where angular jitter is strongly suppressed, and obtain an angular noise-floor of $\approx 350~\text{nanodegrees}$ equivalent to a radial displacement of 3~picometers. Such measurement sensitivity is comparable to the value reported for non-Kramers system with even stronger g-factor anisotropy~\cite{kumar_strongly_2026}, and those achieved with laser interferometry~\cite{hahn_heterodyne_2010, paolino_single_2007}. 

Finally, to qualitatively understand the reduced echo variances at increasing temperature [Fig.~\ref{fig1:invEcho}(d)], we note again that at timescales of 10~ms, dominant spin relaxation in our inversion recovery sequences happens via spin diffusion such that excitations, driven by mutual flip-flop of erbium spins~\cite{bottger_optical_2006, rancic_electron-spin_2022-1}, leave the detection volume or bandwidth of the resonator before they can relax to phonons. Even with modest increase of temperature, their diffusion rate following $\left [ \text{sech} \left(\frac{\hbar \omega_0}{2 k_\text{B} T} \right) \right ]^2$ becomes quite strong, resulting in spatial and spectral smearing of the spectral hole.

\section{Conclusions}
In conclusion, we have directly observed the influence of pulse-tube vibrations on the angular jitter of the spin quantization axis in anisotropic Kramers ions measured in a dilution refrigerator. Although the implications of this study are general, they are particularly relevant to the material system investigated here, $\Er$ in CaWO$_4$, which is considered one of the most promising platforms for quantum memories~\cite{hedges_efficient_2010,sinclair_spectral_2014,businger_optical_2020} and quantum interfacing, owing to its optical transition in the telecom band~\cite{sangouard_quantum_2011, rochman_microwave--optical_2023}. Our findings thus highlight the importance of exploring isotropic crystal hosts, such as $\Er$ in CeO$_2$~\cite{zhang_optical_2024, mistri_clock-transition-enabled_2026}, as potential alternatives for use in quantum-technologies. Additionally, mitigation strategies for vibration damping, either active or passive, can be pursued in future studies.   

%####### Acknowledgements
\section{Acknowledgment}
We thank P. Bertet, A. Doll and M. Rancic for fruitful discussions. This work is supported by Department of Atomic Energy, Government of India, under Project Identification No. RTI 40007, and the Department of science and technology (DST) India for the SERB startup grant SRG/2023/001999. The resonator sample was fabricated at Myfab Chalmers. SDG acknowledges support from the UK Department for Science Innovation and Technology through the UK National Quantum Technologies Programme. The data that support the findings of this study are available from the corresponding author upon reasonable request.

%%###################################################
\renewcommand{\thefigure}{A\arabic{figure}}
\setcounter{figure}{0}
\renewcommand{\theequation}{A\arabic{equation}}
\setcounter{equation}{0}

\appendix
\section{Finding variance} \label{app:a}
When a square $\pi$ pulse is applied on the spins, we are assuming a gaussian spectral hole with bandwidth given by the pulse duration of the $\pi$ pulse,
\begin{align}
H(\delta \omega, ~\Omega) \propto  \exp \left(- \frac{(\Omega- \delta \omega)^2}{2 \sigma^2_H}  \right)
\end{align}
Similarly, the detection window is a gaussian with the same bandwidth as the spectral hole.
\begin{align}
    D(\Omega) \propto  \exp \left(- \frac{\Omega^2}{2 \sigma^2_D} \right)
\end{align}
The detected signal will have a filter function which is a convolusion of the two gaussians
\begin{align}
    F(\delta \omega) &= \int_{-\infty}^{+\infty} H(\delta \omega, \Omega) D(\Omega) d\Omega \nonumber \\
    &= \exp \left(-\frac{\delta \omega^2}{2 \left(\sigma^2_H+\sigma^2_D \right)} \right)
    \label{eq:filter}
\end{align}
with $\sigma_\text{eff}=\sqrt{\sigma^2_H+\sigma^2_D}=\sqrt{2}\sigma_H$, the various constants are absorbed into $K$ in Eq~\eqref{eq:signal}.

To explain the mechanical vibrations, Figure~\ref{fig1:Schematics}(c) shows a combination of harmonics, broad mechanical resonances and a 1/f trend. Since the repetitions of the three-pulse measurement and the pulse tube action are not synced, each single-shot measurement samples this complex waveform at a random point of time, thus the resulting angular dispacement behaves as a stochastic variable and by the Central limit theorem, the probability distribution of a certain angular displacement $\delta \theta$ is given by a Gaussian. Considering a linear relation between detuning $\delta \omega$ and the angular displacement $\delta \theta$, we can write the probability of detuning as
\begin{align}
    P(\delta \omega) = \frac{1}{\sqrt{2 \pi}\sigma_{\omega}} \exp \left(- \frac{\delta \omega^2}{2\sigma^2_{\omega}} \right)
    \label{eq:detune}
\end{align}
with $\sigma_{\omega} = 2 \pi \left| \frac{\partial \nu}{\partial \theta} \right| \sigma_{\theta}$. Now, the variance of the observed signal is calculated as
\begin{align}
    \begin{split}
    \langle (\Delta S)^2 \rangle 
    = \int_{-\infty}^{+\infty} S&^2(\delta \omega) P(\delta\omega) ~d \delta \omega \\
    &- \left[\int_{-\infty}^{+\infty} S(\delta \omega) P(\delta \omega) ~d \delta \omega \right]^2
    \end{split}
\end{align}
We get Eq~\eqref{vareq} as the final result. The constant $K$ is found using the most probable echo magnitude value, since $\delta \omega =0$ is the most probable detuning from Eq~\eqref{eq:detune}, the observed signal will be $S(\delta \omega = 0) = A(1-K)$.

%% Potential referees: Steven Touzard, Daniel Oblak

%##########  Bibliography  ######################
%\bibliography{Er_vibration}


\end{document}